\documentstyle[prl,aps,multicol]{revtex}
\def\narrowtext{}

\begin{document}
\narrowtext
\begin{multicols}{2}

{\bf Giamarchi and Le Doussal Reply}: 
the Comment \cite{balents_comment} reexamines some
of the physics of the moving glass introduced in \cite{giamarchi_prl}.
In \cite{giamarchi_prl} we demonstrated that
a periodic structure driven along $x$ on a disordered substrate
experiences a transverse static pinning force $F_y^{\rm stat}$ along
$y$. A key point (formula (2) of\cite{giamarchi_prl}) is that this force,
missed in previous studies, originates {\it only} from the periodicity along $y$
and the uniform density modes along $x$, i.e the smectic-like modes. It
leads to novel glassy effects, the main predictions
being the existence of a transverse critical force $F^y_c$ and of static pinned
channels of motion, subsequently observed in
numerical simulations \cite{moon_mglass}.

\cite{balents_comment} claims the following: (i) in addition to $F_y^{\rm
stat}$ there is also a $u$-independent random force $f_d(r)$ generated
in the direction of motion (ii) due to $f_d(r)$
the elastic theory has to be reconsidered and correlation
functions have to be recalculated.
(iii) dislocation unbinding will occur for $d \le 3$.
Let us answer specifically to each point.

We do not disagree with (i), i.e that there is an effective random force
along $x$, though this rather subtle question cannot be
settled by naive perturbation theory alone. In particular
the argument of time translational invariance of \cite{balents_comment} leaves the possibility
of $u_y$ dependence and does {\it not}
prove that the random force is $u$-independent. The existence of such a random force
needs therefore to be proven carefully, through e.g. an RG calculation [3].
Such a calculation also yields $F^y_c$ and other important physical effects.
In particular we find
\cite{mglass_long} that a random force is generated along $y$, a point unnoticed
in \cite{balents_comment}.

Concerning the consequences of such a random force $f_d(r)$ along $x$
within the elastic theory, we disagree
with (ii). As mentionned above, the properties of the moving glass
rely only on the periodicity along $y$ and are thus,
to a large extent, independent of the
details of the structure along $x$, i.e the behaviour
of $u_x$. This is illustrated by
the fact that the authors of \cite{balents_comment} find ``surprisingly''
that $f_d(r)$ does not
change the transverse correlator $B_y(r)$ obtained in \cite{giamarchi_prl}
(note that our result is incorrectly quoted since
we find a logarithm only in $d=3$).
This is a simple consequence
of the related fact that the compression modes are responsible for
the moving glass \cite{giamarchi_prl}.
In fact setting {\it formally} $u_x=0$ leads to the useful equation
(3) of \cite{giamarchi_prl} describing the transverse physics
of the moving glass.

Concerning point (iii), as in the statics
\cite{giamarchi_statics}
the elastic description is only a starting point,
and the issue of whether dislocations 
are generated is a well-known difficult problem.
Only a controlled calculation including dislocations could settle
this point, but it has not yet been
performed. However, qualitative arguments \cite{mglass_long}
indicate that complete topological
order exists at weak disorder or large velocity in $d=3$ \cite{giamarchi_prl}
(at variance with \cite{balents_comment}) while it disappears in $d=2$.

More importantly, our picture of pinned channels should
remain valid even with dislocations, 
as long as periodicity along $y$ is maintained, i.e
for a smectic-like structure.
Before the channel picture was identified in \cite{giamarchi_prl},
it was unclear how dislocations affect the moving structure.
The existence of channels then
{\it naturally suggests a scenario} \cite{mglass_long} by which dislocations will
appear:
When the periodicity along $x$ is retained, e.g presumably in $d=3$ at
weak disorder, the channels are coupled along $x$. Upon increasing disorder
or decreasing velocity in $d=3$, or in $d=2$,
decoupling between channels can occur,
reminiscent of static decoupling in a layered geometry
\cite{decoupling}.
Dislocations are then inserted between the layers,
naturally leading to a ``flowing smectic'' glassy state,
recently observed in numerical simulations in $d=2$
\cite{moon_mglass}. 
Indeed the transverse smectic order is likely to be more stable
that topological order along $x$, because of particle conservation \cite{mglass_long}.
Decoupling due to the random force
advocated in \cite{balents_comment}, may be one realization of this 
general scenario. In fact [2] naturally suggests that 
transitions from elastic to plastic flow may be studied
as ordering transitions in the structure of channels.

In Conclusion, since only periodicity in the
{\it transverse} direction is needed and contrarily to the
claim of \cite{balents_comment},
\cite{giamarchi_prl} describes correctly the main (transverse)
physics of the moving glass and provides the correct starting
point to study \cite{balents_comment,mglass_long} interesting physical extensions
such as behaviour of $u_x$, random forces, additional linear and non linear
terms coming from anharmonicity, channel
decoupling and dislocations.

\bigskip

T. Giamarchi

Laboratoire de Physique des Solides, Universit{\'e} Paris--Sud,
B{\^a}t. 510, 91405 Orsay, France.

Pierre Le Doussal

CNRS-Laboratoire de Physique Th{\'e}orique
de l'Ecole Normale Sup{\'e}rieure, 24 Rue Lhomond,
F-75231 Paris Cedex 05, France

\end{multicols}
\end{document}